\documentclass{article}
\usepackage{amsmath}
\usepackage{graphicx}
\title{The colored flux tube}
\author{Vladimir Dzhunushaliev
\thanks{
E-mail: dzhun@hotmail.kg}}
\date{}
\begin{document}
\maketitle

\begin{center}
\textit{
Institut f\"ur Mathematik,  
Universit\"at Potsdam, D-14469, Potsdam, Germany\\
and \\
Dept. Phys. and Microel. Engineer., Kyrgyz-Russian
Slavic University\\
Bishkek, Kievskaya Str. 44, 720021, Kyrgyz
Republic}
\end{center}

\begin{abstract}
It is shown that in the SU(2) Yang - Mills - Higgs theory with broken 
gauge symmetry a flux tube solution filled with a color longitudinal  
electric field exists. The origin of the gauge symmetry breakdown 
for this case is discussed.
\end{abstract}

\section{Introduction}

The quantum chromodynamics (QCD) describes the forces between quarks which are 
elementary blocks of hadrons. In contrast with quantum electrodynamics 
QCD has a very strong interaction between quanta of gauge fields. 
Probably the consequences of such nonlinearity is the appearance of a flux 
tube filled with a longitudinal color electric field between interacting quark and antiquark. The existence of such tube leads to a confinement of quarks: 
quark and antiquark are connected with the flux tube which gives a 
constant force between these particles. The quarks are the sources 
of a color electric field filling the flux tube. Therefore the derivation of 
the flux tube solutions in QCD is very important problem. Evidently this 
problem is quantum not classical one as many years of the attempts to obtain such 
solutions in the classical gauge theories have not any success. Altough it is 
well known that in dual theories exist so-called Nielsen-Olesen flux tube solutions \cite{no} which are filled with an Abelian magnetic field and BPS strings 
\cite{kneipp}. Mathematically these 
solutions are identical to the flux tubes in a superconductor where the magnetic 
field is 
pushed out from the superconductor. Such correspondence between color flux tube 
in QCD and magnetic flux tube in superconductor allows us to suppose 
that the QCD flux tube in some sense is a dual picture of a Meissner 
effect in the superconductivity \cite{tHooft75}, \cite{mandelstam} : 
magnetic field is changed on electric one and a condensate of Cooper pairs on 
a condensate of (probably) magnetic monopoles. 
\par 
In Ref. \cite{dzhun1} it was shown that in the Euclidean spacetime there is 
a flux tube solution (colored flux tube) of the SU(2) Yang - Mills - Higgs gauge 
theory with broken gauge symmetry. The derived solution is filled with a color longitudinal SU(2) electric field. The analysis shows that the symmetry breakdown is necessary for the existence of colored flux tube as this solution is regular only 
for a discrete spectrum of mass $m \neq 0$ which describes the gauge symmetry 
breakdown. In this letter we would like to show that the same mechanism is in 
the action for the Lorentzian spacetime, too. 

\section{Flux tube equations}

We will start from the SU(2) Yang - Mills - Higgs field equations 
with broken gauge symmetry
\begin{eqnarray}
  \mathcal{D}_\nu F^{a\mu\nu} &=& g \epsilon^{abc} \phi^b 
  \mathcal{D}^\mu \phi^c - \left( m^2 \right)^{ab} A^{b\mu} , 
\label{sec1-10}\\
  \mathcal{D}_\mu \mathcal{D}^\mu \phi^a &=& -\lambda \phi^a 
  \left(
  \phi^b \phi^b - \phi^2_\infty
  \right) 
\label{sec1-20}
\end{eqnarray}
here $F_{a\mu\nu} = \partial_\mu A^a_\nu - \partial_\nu A^a_\mu 
+ g \epsilon^{abc} A^b_\mu A^c_\nu$ is the field tensor for the SU(2) gauge 
potential $A^a_\mu$; $a,b,c = 1,2,3$ are the color indices; 
$D_\nu [\cdots ]^a = \partial_\nu [\cdots ]^a + 
g \epsilon^{abc} A^b_\mu  [\cdots ]^c$ is the gauge derivative; $\phi^a$ 
is the Higgs field; $\lambda , g$ and $\phi_\infty$ are some constants; 
$\left( m^2 \right)^{ab}$ is a masses matrix which 
destroys the gauge invariance of the Yang - Mills - Higgs theory, 
here we choose $\left( m^2 \right)^{ab} = diag \left\{ m^2_1, m^2_2, 0 \right\}$. 
\par
The solution we search in the following form 
\begin{equation}
    A^1_t(\rho) = \frac{f(\rho)}{g} ; \quad A^2_z(\rho) = \frac{v(\rho)}{g} ; 
    \quad \phi^3(\rho) = \frac{\phi(\rho)}{g} 
\label{sec1-25}
\end{equation}
here $z, \rho , \varphi$ are cylindrical coordinate system. The substitution 
into the Yang - Mills - Higgs equations \eqref{sec1-10} \eqref{sec1-20} gives us 
\begin{eqnarray}
    f'' + \frac{f'}{x} &=& f \left( \phi^2 + v^2 - m^2_1 \right),
\label{sec1-30}\\
    v'' + \frac{v'}{x} &=& v \left( \phi^2 - f^2 - m^2_2 \right),
\label{sec1-40}\\  
    \phi'' + \frac{\phi'}{x} &=& \phi \left[ - f^2 + v^2 
    + \lambda \left( \phi^2 - \phi^2_\infty \right)\right]
\label{sec1-50}
\end{eqnarray}
here we redefined $\phi /\alpha \rightarrow \phi$, $f /\alpha  \rightarrow f$, 
$v /\alpha  \rightarrow v$, $\phi_\infty /\alpha \rightarrow \phi_\infty$, 
$m_{1,2} /\alpha  \rightarrow m_{1,2}$, $\rho \alpha  \rightarrow x$; the 
constant $\alpha$ will be defined later. 
The similar equations with the presence of $A^a_\varphi$ and 
without masses $m_{1,2}$ was investigated in Ref's \cite{Obukhov:1996ry} 
\cite{Dzhunushaliev:1999fy} where it was shown that the corresponding solutions 
becomes singular either on a finite distance from the axis $\rho = 0$ or 
on the infinity. The solutions with the masses $m_{1,2}$ and a magnetic 
field $H_z$ were obtained in Ref. \cite{dzhun} and the result is that 
the following versions of the flux tube 
exist: (1) the flux tube filled with electric/magnetic fields 
on the background of an external constant magnetic/electric field; (2) 
the Nielsen-Olesen flux tube dressed with transversal color electric 
and magnetic fields. 

\section{Numerical algorithm}

We will solve the equations set \eqref{sec1-30}-\eqref{sec1-50} with an 
iterative procedure which is described in Ref. \cite{dzhun1}. 
Eq. \eqref{sec1-40} on the $i$ step has the following form 
\begin{equation}
    v_i'' + \frac{v_i'}{x} = v_i \left( \phi^2_{i-1} - f^2_{i-1} - m^2_{2,i} \right),
\label{sec2-10}\\
\end{equation}
here the functions $f_{i-1}$ and $\phi_{i-1}$ was defined on $(i-1)$ step and the 
null approximation for the functions $f(x)$ and $\phi(x)$ 
is $\phi_0(x) = 1.3 - 0.3/\cosh^2 (x/4)$ and 
$f_0(x) = 0.2/\cosh^2(x)$. The numerical investigation for Eq. 
\eqref{sec2-10} shows that there is a number $m^*_{2,i}$ for which: for 
$m_{2,i} > m^*_{2,i}$ the solution is singular $v_i(x) \rightarrow - \infty$ 
by $x \rightarrow \infty$ and for $m_{2,i} < m^*_{2,i}$ the solution is also singular 
$v_i(x) \rightarrow + \infty$ by $x \rightarrow \infty$. It means 
that there is such number $m_{2,i} = m^*_{2,i}$ for which the solution 
$v_i(x)$ is regular one. The number $m^*_{2,i}$ is defined with the method of 
the iterative approximation. Here is 
necessary that the functions $f_{i-1}(x) \rightarrow 0$ and 
$\phi_{i-1}(x) \rightarrow \phi^*_{\infty, i-1}$ by 
$x \rightarrow \infty$. 
\par 
Eq. \eqref{sec1-30} on this step has the following form 
\begin{equation}
    f_i'' + \frac{f_i'}{x} = f_i \left( \phi^2_{i-1} + v^2_i - m^2_{1,i} \right),
\label{sec2-15}\\
\end{equation}
here the function $\phi_{i-1}$ was defined on $(i-1)$ step and the 
function $v_i$ is the solution of Eq. \eqref{sec2-10}. The numerical investigation for Eq. 
\eqref{sec2-15} shows that there is a number $m^*_{1,i}$ for which: for 
$m_{1,i} > m^*_{1,i}$ the solution is singular $f_i(x) \rightarrow - \infty$ 
by $x \rightarrow \infty$ and for $m_{1,i} < m^*_{1,i}$ the solution 
is also singular $f_i(x) \rightarrow + \infty$ by $x \rightarrow \infty$. It means 
that exists such number $m_{1,i} = m^*_{1,i}$ for which the solution 
$f_i(x)$ is regular one. The number 
$m^*_{1,i}$ is defined with the method of iterative approximation. It is 
necessary that the functions $v_i(x) \rightarrow 0$ and 
$\phi_{i-1}(x) \rightarrow \phi^*_{\infty,i}$ by 
$x \rightarrow \infty$. 
\par
The next step on the $i$ iteration is solving of Eq. \eqref{sec1-50}
\begin{equation}
    \phi_i'' + \frac{\phi_i'}{x} = \phi_i \left[ -f^2_i + v^2_i + 
    \lambda \left( \phi^2_i - \phi^2_{\infty ,i} \right)\right]
\label{sec2-20}
\end{equation}
here the functions $f_i$ and $v_i$ are the solutions of Eq's 
\eqref{sec2-10} \eqref{sec2-15}. 
The numerical investigation for this equation shows that there is a number 
$\phi^*_{\infty ,i}$ for which: for $\phi_{\infty ,i} > \phi^*_{\infty ,i}$ 
the function $\phi_i (x)$ oscillates with decreasing amplitude, for 
$\phi_{\infty ,i} < \phi^*_{\infty ,i}$ the function 
$\phi_i (x) \rightarrow + \infty$ by $x \rightarrow x_0$. The number 
$\phi^*_{\infty ,i}$ defines a regular solution $\phi_i(x)$. It is 
necessary that the functions $v_i(x) \rightarrow 0$ and 
$f_i(x) \rightarrow 0$ by $x \rightarrow \infty$. 
\par
The value $f(0)$ can be arbitrary but we choose them by such a way that 
$f(0)/\alpha = 0.2$, the other initial conditions are $v(0) = 0.5$ 
and $\phi(0) = 1.0$. Thus in the equations set \eqref{sec1-30}-\eqref{sec1-50} 
we have three independent parameters $\lambda, v(0)$ and $\phi(0)$. In the 
calculations presented here we take $\lambda = 0.1$. 
\par
\begin{figure}[h]
    \begin{center}
    \fbox{
        \includegraphics[width=10cm,height=7cm]{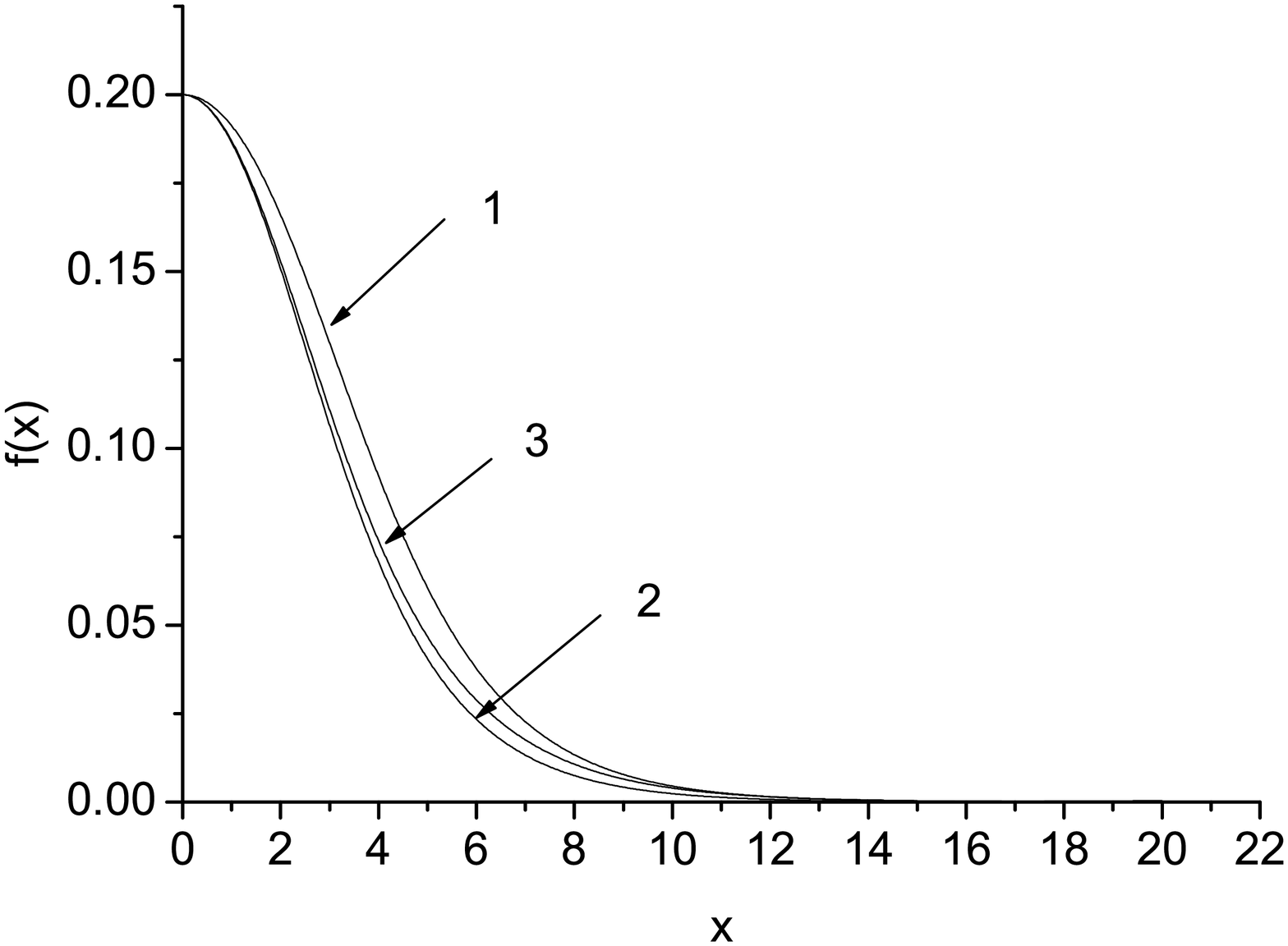}}
        \caption{The first three iteration for the $A^1_t$ gauge potential component.}
        \label{fig1}
    \end{center}
\end{figure}
\begin{figure}[h]
    \begin{center}
    \fbox{
        \includegraphics[width=10cm,height=7cm]{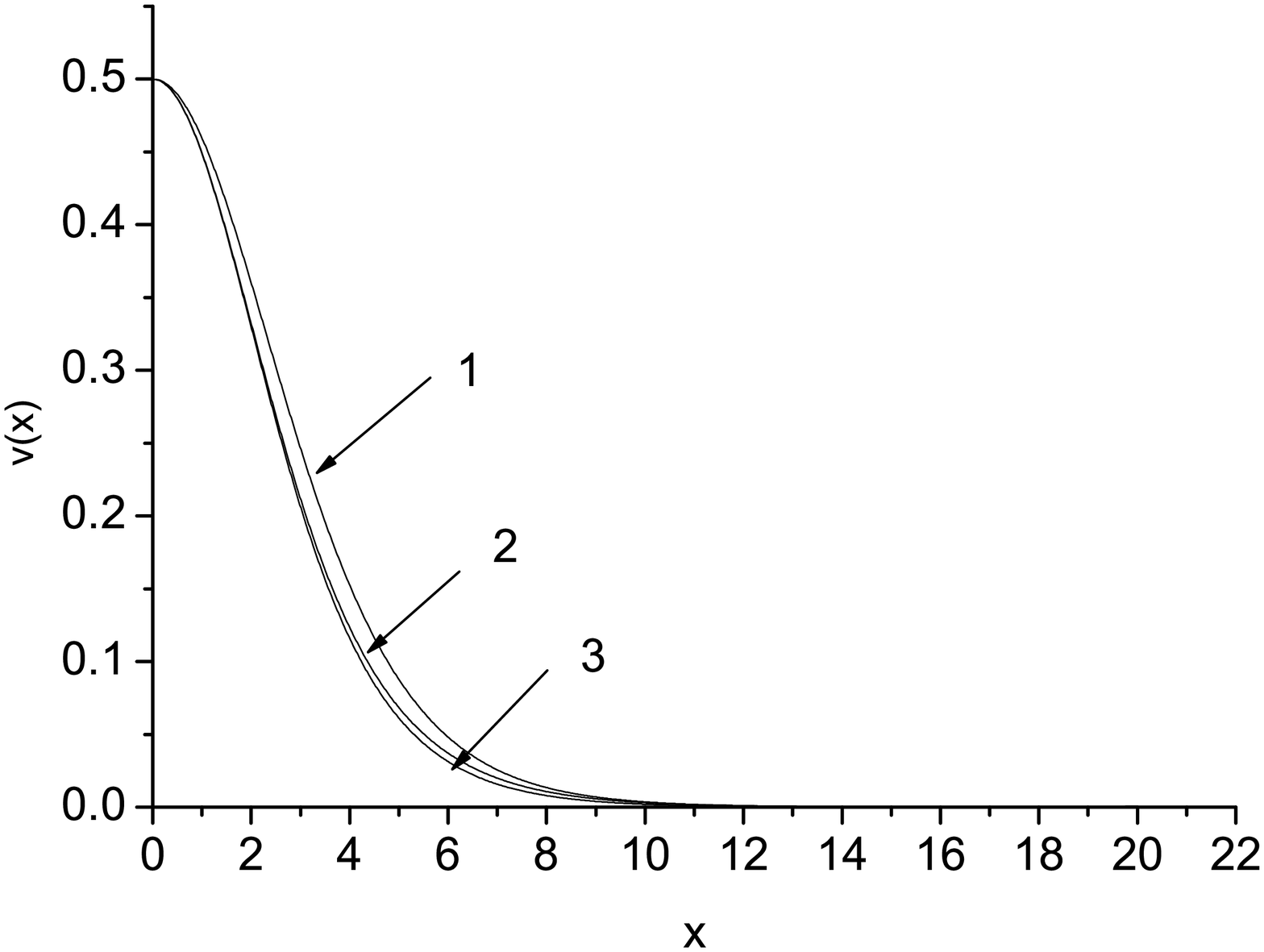}}
        \caption{The first three iteration for the $A^2_z$ gauge potential component.}
        \label{fig2}
    \end{center}
\end{figure}
\begin{figure}[h]
    \begin{center}
    \fbox{
        \includegraphics[width=10cm,height=7cm]{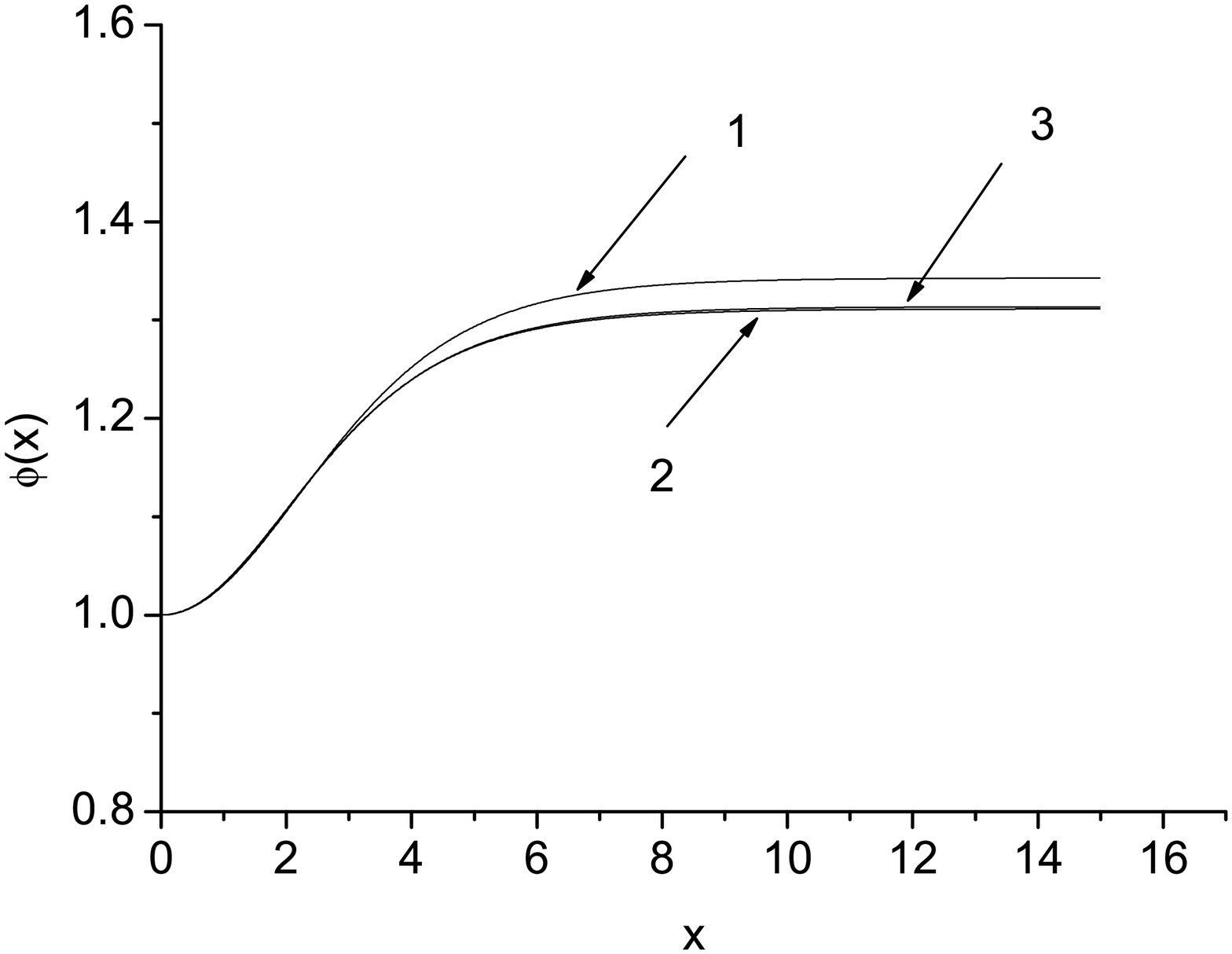}}
    \caption{The first three iteration for the $\phi^3$ scalar field.}
    \label{fig3a}
    \end{center}
\end{figure}

The iterative process described above gives us the $m^*_{(1,2), i}$ and 
$\phi^*_{\infty ,i}$ presented on Table \ref{table1}. 
\begin{table}[h]
    \begin{center}
        \begin{tabular}{|c|c|c|c|}\hline
          i & 1 & 2& 3 \\ \hline
            $m^*_{1,i}$ & 1.1944813\ldots & 1.2362642\ldots & 1.232607\ldots \\ \hline
            $m^*_{2,i}$ & 1.1405457\ldots & 1.1815878\ldots & 1.181085\ldots \\ \hline
            $\phi^*_{\infty ,i}$ & 1.3426585\ldots & 1.3112676\ldots & 1.313513\ldots \\ \hline
        \end{tabular}
    \end{center}
    \caption{The iterative parameters $m^*_{(1,2), i}$ and $\phi^*_{\infty ,i}$.}
    \label{table1}
\end{table}
The functions $f_i(x), v_i(x)$ and $\phi_i(x)$ for $i=1,2,3$ are presented 
on Fig's \ref{fig1}, \ref{fig2} and \ref{fig3a}. 

\begin{figure}[h]
    \begin{center}
    \fbox{
        \includegraphics[width=10cm,height=7cm]{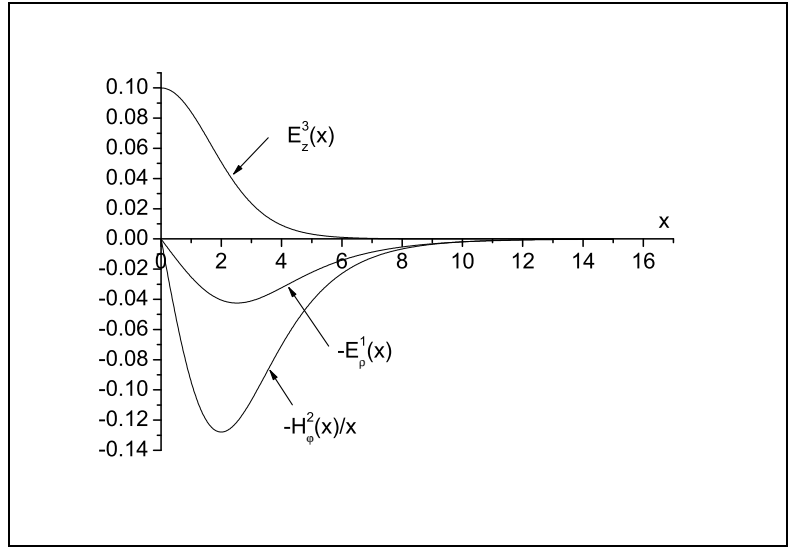}}
    \caption{The color electric $E^3_z(x)$, $E^1_\rho(x)$ and magnetic 
    $H^2_\varphi (x)$ fields} 
    \label{fig4}
    \end{center}
\end{figure}

\par 
By the definition the color electric and magnetic fields are 
\begin{eqnarray}
    E^3_z(x) &=& F^3_{tz} = \frac{f(x) v(x)}{g}  ,
\label{sec2-30}\\
  E^1_\rho(x) &=& F^1_{t \rho} = -\frac{f'(x)}{g} ,
\label{sec2-40}\\
  H^2_\varphi (x) &=& x \epsilon_{\varphi \rho z} F^{2\rho z} = 
  -x \frac{v'(x)}{g}.
\label{sec2-45}
\end{eqnarray}
These fields are presented on Fig. \ref{fig4}. 
\par 
From Eq's \eqref{sec1-30}-\eqref{sec1-50} it is easy to see that the 
symptotical behaviour of the regular solutions $f^*(x)$, $v^*(x)$ and $\phi^*(x)$ is 
\begin{eqnarray}
    f^*(x) &=& f_0 \frac{e^{-x\sqrt{\phi^{*2}_\infty - m_1^{*2}}}}{\sqrt{x}} + 
    \cdots ,
\label{sec2-50}\\
    v^*(x) &=& v_0 \frac{e^{-x\sqrt{\phi^{*2}_\infty - m_2^{*2}}}}{\sqrt{x}} + 
    \cdots ,
\label{sec2-60}\\
  \phi^*(x) &=& \phi^*_\infty - 
  \phi_0 \frac{e^{-x\sqrt{2\lambda \phi^{*2}_\infty}}}{\sqrt{x}} + \cdots 
\label{sec2-70}
\end{eqnarray}
where $f_0, v_0$ and $\phi_0$ are some constants. 
\par
\begin{figure}[h]
    \begin{center}
    \fbox{
        \includegraphics[width=10cm,height=7cm]{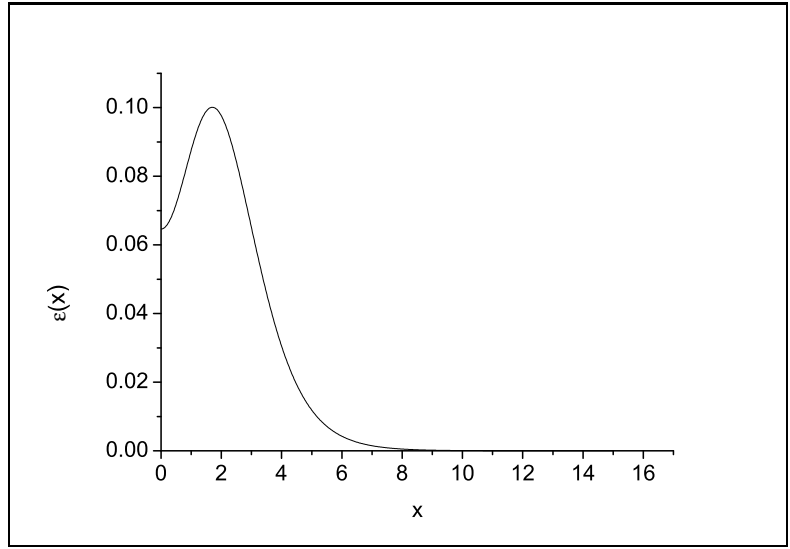}}
    \caption{The energy density $\epsilon(x)$.} 
    \label{fig5}
    \end{center}
\end{figure}

The energy density is 
\begin{eqnarray}
    2 g ^2 \epsilon(x) = {f'}^2(x) + {v'}^2(x) + {\phi '}^2(x) + 
    v^2(x) f^2(x) + v^2(x) \phi^2(x) + f^2(x) \phi^2(x) + 
\nonumber \\
    m^*_1 f^2(x) - m^*_2 v^2(x) + \frac{\lambda}{2} 
    \left( \phi^2 - \phi^2_\infty \right)^2 
\label{sec2-80}
\end{eqnarray}
and presented in Fig. \ref{fig5}. The linear energy density will be finite as 
all terms in \eqref{sec2-80} have exponential decreasing at the infinity.
\par 
The flux of the longitudinal electric field is 
\begin{equation}
    \Phi = \int E^3_z ds = 2 \pi \int^{\infty}_{0} \rho \frac{f(\rho)v(\rho)}{g} 
    d\rho = \frac{2 \pi}{g} \int^{\infty}_{0} x f(x) v(x) dx < \infty.
\label{sec2-90}
\end{equation}
We see that the flux $\Phi$ depends on the parameters $\lambda, v(0)$ 
and $\phi (0)$ as the solution $f(x)$ and $v(x)$ depend on these 
parameters which is in contrast with such topological solutions as 
't Hooft-Ployakov monopole and Nielsen-Olesen flux tube. It is the 
indication of the fact that the colored flux tube solution is not 
the consequence of any differential equations of the first order as it 
is for the topological solutions. In other words the colored flux tube 
is a dynamical object not topological one. 

\section{Colored flux tube as a pure quantum object}

In Ref. \cite{dzhun1} was presented some arguments that the obtained 
there colored flux tube has a pure quantum origin. 
The arguments are that the SU(3) quantum Yang - Mills theory 
in some approximation can be reduced to the classical SU(2) Yang - Mills - Higgs 
theory plus some extra term which is zero for the ansatz \eqref{sec1-25}. 
We will describe this $SU(3) \rightarrow SU(2)$ reduction following to 
Ref. \cite{vdsin2}. At first the SU(3) gauge potential on ordered 
and disordered phases should be decomposed: 
\begin{enumerate}
\item The gauge field components $A^a_\mu \in SU(2), a=1,2,3$ 
      belonging to the small subgroup $SU(2) \subset SU(3)$ are in an ordered 
      phase:
\begin{equation}
  \left\langle A^a_\mu (x) \right\rangle  \approx (A^a _{\mu} (x))_{cl}.
\label{sec3-10}
\end{equation}
      The subscript $(\cdots)_{cl}$ means that the SU(2) components of the SU(3) 
      gauge field can be considered as almost a classical field. 
      $\left\langle \ldots \right\rangle$ is a quantum average. 
\item The gauge field components $A^m_\mu$ (m=4,5, ... , 8) and 
      $A^m_\mu \in SU(3)/SU(2)$) belonging to the coset SU(3)/SU(2) are in 
      a disordered phase. i.e. they are a condensate of the coset components 
      of the SU(3) gauge field. It means that 
\begin{equation}
  \left\langle A^m_\mu (x) \right\rangle = 0, 
  \quad \text{but} \quad 
  \left\langle A^m_\mu (x) A^n_\nu (x) \right\rangle \neq 0 .
\label{sec3-20}
\end{equation}
      These degrees of freedom are pure quantum degrees and are involved in the 
      equations for the ordered phase as an averaged field distribution of coset 
      components. 
\end{enumerate}
In order to simplify the 2 and 4-points Green's functions the following assumptions and simplifications was made:
\begin{enumerate}
\item 
    The correlation between coset components $A^m_\mu (y)$ and $A^n_\nu (x)$ 
    in two points $x^\mu$ and $y^\mu$ is 
    \begin{equation}
        \left\langle A^m_\mu (y) A^n_\nu (x) \right\rangle =
        - \frac{1}{3}f^{mpb} f^{npc} \eta_{\mu \nu} \phi^b (y) \phi^c (x) .
    \label{sec3-30}
    \end{equation}
    where $f^{abc}$ is the structural constants of the SU(3) group. 
    Here we have to indicate that it is correct only for static fields, 
    i.e. $x^0 = y^0$. 
\item There is not any correlation between 
      ordered (classical) and disordered (quantum) phases 
    \begin{equation}
      \left\langle f(a^a_\mu) g(A^m_\nu) \right\rangle =
      f(a^a_\mu)  \left\langle g(A^m_\mu) \right\rangle
    \label{sec3-50}
    \end{equation}
    where $f$ and $g$ are arbitrary functions. 
\item 
    The 4-point Green's function can be approximated as  
    \begin{eqnarray}
        \left\langle
        A^m_\alpha (x) A^n_\beta (y) A^p_\mu (z) A^q_\nu (u) 
        \right\rangle = 
        \nonumber \\
        \left(
        E^{mnpq}_{1,abcd} \eta_{\alpha\beta} \eta_{\mu\nu} + 
        E^{mpnq}_{2,abcd} \eta_{\alpha\mu} \eta_{\beta\nu} + 
        E^{mqnp}_{3,abcd} \eta_{\alpha\nu} \eta_{\beta\mu}
        \right) 
        \phi^a (x) \phi^b(y) \phi^c (z) \phi^d(u) 
\label{sec3-60}
\end{eqnarray}
here $E^{mnpq}_{1,abcd}, E^{mpnq}_{2,abcd}, E^{mqnp}_{3,abcd}$ are some 
constants and this approximation is also valid for the static fields. 
\end{enumerate} 
Eq's \eqref{sec3-30} and \eqref{sec3-60} tell us that this approach can be called 
``one function approximation``.
\par 
The final result of Ref.\cite{vdsin2} is that the initial SU(3) Lagrangian 
\begin{equation}
    \mathcal{L}_{SU(3)} = -\frac{1}{4}F^A_{\mu\nu} F^{A\mu\nu}, \, 
    A = 1,2, \cdots 8 
\label{sec3-70}
\end{equation}
with these assumptions and simplifications can be reduced to the 
SU(2) Yang - Mills - Higgs Lagrangian 
\begin{equation}
  \mathcal{L}_{SU(2)} = - \frac{1}{4}  F^a_{\mu\nu} F^{a\mu\nu} + 
  \frac{1}{2} \left(
  \partial_\mu \phi^a + \frac{g}{2} \epsilon^{abc} A^b_\mu \phi^c
  \right)^2 + \frac{m^2_\phi}{2} (\phi ^a \phi ^a ) 
  - \lambda \left( \phi^a \phi^a \right)^2  +
  \frac{g^2}{2} a_{\mu} ^b \phi ^b a^{c \mu} \phi ^c .
\label{sec3-80}
\end{equation}
It is necessary to note that the term $\frac{m^2_\phi}{2} (\phi ^a \phi ^a )$ 
here also presents the assumed gauge symmetry breakdown. 
We see that the first term $F^a_{\mu\nu} F^{a\mu\nu}$ is the SU(2) Lagrangian 
for the ordered phase $A^a_\mu$ and the Higgs Lagrangian is presented 
with the next three terms. Let us note that there is an additional gauge noninvariant term 
$\frac{g^2}{2} a_{\mu} ^b \phi ^b a^{c \mu} \phi ^c$ but for the ans\"atz 
\eqref{sec1-25} the corresponding terms in field equations are zero. 
\par
Finally, in the context of offered here $SU(3) \rightarrow SU(2)$ reduction 
the colored flux tube obtained above is \textit{a pure quantum object in the 
SU(3) Yang - Mills theory}. 

\section{Discussion}

The main result of this letter is that the confinement problem in QCD 
and the gauge symmetry breakdown can be closely connected. 
In our calculations the symmetry 
breaking term is presented by the mass term $(m^2)^{ab} A^b_\mu$. Of course 
this term is inserted by hand. The problem for the derivation of this term 
from the first principles is the absence of non-perturbative technique for 
the calculations in quantum field theories with strong interactions. 
Although some calculations \cite{dudal} show us that such terms can be 
obtained in quantum gauge theories on the perturbative level. 
One can indicate the probable connection of our problem with 
a Coleman - Weinberg mechanism \cite{coleman} in 
$\lambda \phi ^4$-theory. This mechanism gives us an additional term 
which changes an initial potential term and makes from its something like 
a Higgs potential (Mexico hat). The origin of this phenomenon is the 
presence of the nonlinear potential $\lambda \phi^4$ in the theory. One can 
suppose that the similar mechanism will work in the quantum gauge theories 
where there is the potential term like to $(A)^4$, but of course that 
here we have to use a non-perturbative technique which probably 
will be similar to the quantization method applied by Heisenberg to a 
non-linear spinor field \cite{heis}. 
\par 
Another interesting property of the colored flux tube solution is 
the discreteness of masses $m^*_{1,2}$ and $\phi^*_\infty$ which gives 
the finitiness of the corresponding solution. We can make an assumption 
that it is connected with the fact that these parameters are quantum 
corrections and as it takes place in any quantum theory something 
should be quantized. 
\par
In conclusion we would like to note that the derived here the 
longitudinal electric filed is essentially nonlinear: it is not 
the gradient of some function but it appears from the nonlinear term 
$g \epsilon^{abc} A^b_\mu A^c_\nu$ of the non-Abelian tensor 
$F^a_{\mu\nu}$, i.e. the Maxwell eletrodynamics can not have such 
flux tube solutions even with broken gauge symmetry. One can say that 
the nonlinearity and symmetry breakdown of non-Abelian gauge theories 
probably are the key ingredients of confinement problem in QCD. 

\section{Acknowledgments}
I am very grateful to the Alexander von Humboldt foundation for the 
financial support of this work and H.-J- Schmidt for the invitation to 
research in Potsdam University.

\end{document}